# Application of Multicore Optical Fibers in Astronomy

Nemanja Jovanovic[1,2], Olivier Guyon[1,3,4,5], Hajime Kawahara[6,7], and Takayuki Kotani[8]
(1) Subaru Telescope, National Astronomical Observatory of Japan, National Institutes of Natural Sciences (NINS),
650 North A'Ohoku Place, Hilo, HI, 96720, U.S.A.
(2) Department of Physics and Astronomy, Macquarie University, NSW 2109, Australia
(3) Steward Observatory, University of Arizona, Tucson, AZ, 85721, U.S.A.
(4) College of Optical Sciences, University of Arizona, Tucson, AZ 85721, U.S.A.
(5) Astrobiology Center of NINS, 2-21-1, Osawa, Mitaka, Tokyo, 181-8588, Japan
(6) Department of Earth and Planetary Science, The University of Tokyo, Tokyo 113-0033, Japan
(7) Research Center for the Early Universe, School of Science, The University of Tokyo, Japan
(8) National Astronomical Observatory of Japan, 2-21-1 Osawa, Mitaka, Japan
*jovanovic.nem@gmail.com*

**Abstract:** Multicore fibers are gaining growing attention in astronomy. The two main attributes which make them attractive for astronomy are that they reduce the distance between cores and hence have a superior fill factor to other approaches and they offer the possibility to transport light in many channels with the overhead of only having to handle a single fiber. These properties are being exploited to realize miniature integral field units that can transport light from different regions in a focal plane to a spectrograph for example. Here we offer an overview of several applications where multicore fibers are now being considered and applied to astronomical observations to enhance scientific yield.

## 1. Introduction

Astronomy has greatly benefitted from the major technical development that photonic technologies have undergone over the past 20 years. Of all photonic technologies, optical fibers have been used most extensively in astronomical instrumentation. The most common application for fibers is for light transport from the focal plane of a telescope to a spectrograph located in a stable, well-isolated room [1]. This simple idea formed the basis for realizing efficient massive (100 000+ target) spectroscopic surveys for the first time. For example the 2 degree Field Galaxy Redshift Survey produced the first map of the structure of the Universe [2,3]. It looked at 200 000+ galaxies which was only possible because of the ability to collect the spectrum of hundreds of targets in a single field simultaneously and the flexibility to rapidly reconfigure the fibers for new fields.

These early multi-object spectroscopic surveys focused on isolated targets within a field that could be addressed by individual optical fibers. However, it quickly became apparent that fibers with more than a single core, or a bundle of fibers could be useful for creating spectroscopic maps across extended targets like galaxies. A device, which segments the image of the target in the focal plane by way of multiple optical fibers, is commonly referred to as an integral field unit (IFU). People quickly realized that a limitation to stacking fibers and forming an efficient bundle was the limited fill factor: the finite cladding of the fibers meant that there is a lot of dead space, which is not collected by the bundle. To some extent some of this limitation can be overcome by placing the fiber bundle behind an array of micro-lenses that have a 100% fill factor. However, micro-optics add complexity, cost and can compromise other performance metrics such as throughput, focal ratio degradation, and so on. For these reasons the field of astronomy began to consider multi-core fibers.

For the purposes of this work we define a multi-core fiber as any fiber, which supports numerous cores, be it single-mode of multi-mode within a single cladding and is formed with traditional techniques which involve tapering (to distinguish from stacking fibers in a tube and gluing them in place). The aim of this paper is to offer an overview of several applications where multi-core fibers are currently being used in astronomy.

## 2. Hexabundles for spatially resolved spectroscopy

The Sydney-AAO Multi-object Integral field spectrograph (SAMI) project aimed at carrying out a large spatially resolved spectroscopic survey of galaxies [4]. As the 3.9-m Anglo-Australian Telescope (AAT) is a seeing-limited telescope (has no adaptive optics), coupling was only possible into multimode fibers (MMF). Typical MMFs considered for the project early on had core diameters of 105 μm and cladding diameters of 125 μm. Although, in linear dimensions the core was 84% of the cladding diameter, in area the cladding was 42% larger. This means that a lot of light was lost in-between the cores of neighboring fibers. For this reason the SAMI project pioneered the process to construct custom multi-core fibers with MMFs with much higher fill factors. The project subsequently used MMFs that had their cladding diameters etched down to 115 μm. 61 such fibers were inserted into a single fluorine-doped capillary and the end was lightly tapered to fuse the fibers without distorting the cores. These devices were dubbed "hexabundles" because of the appearance of their packing. A typical hexabundle is shown in Fig. 1 below. For details about how the hexabundles were optimized for the science case, please refer to [5].

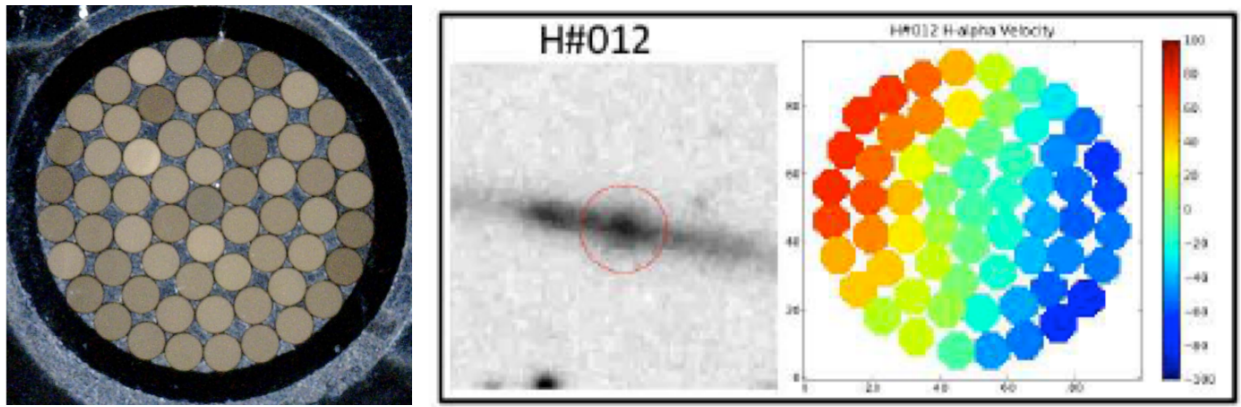

Fig. 1. (Left) An image of a hexabundle. (Middle) An image of a galaxy with the size of the bundle overlaid. (Right) Relative velocity map of a galaxy clearly highlighting rotation of the galaxy. Images taken from S. Croom et al. (2012) [4].

An image of a galaxy observed by the SAMI project is shown in the middle panel of Fig. 1 with an overlay that represents the size of the hexabundle. A velocity map showing the relative motion of different components of that galaxy is shown in the right panel of Fig. 1. This elegantly reveals the rotation of the galaxy as well as the axis about which the galaxy rotates. The SAMI project by virtue of these custom built multi-core fibers has now obtained 700+ similar velocity maps of galaxies and made numerous discoveries from this data.

## 3. Photonic lanterns for efficient conversion of seeing-limited to diffraction-limited light

The photonic lantern is a device developed to channel light from a MMF to a SMF efficiently [6]. Although it is not possible to efficiently direct light from a MMF to a single SMF, it is possible to make this conversion efficient if the number of degrees of freedom (number of modes) of the MMF, are preserved. This is to say that if the light from the MMF were to be split into multiple SMFs, where the number of SMFs was greater than or equal to the number of modes supported by the MMF, then the transition from a MMF to a SMF format could be made efficiently [7]. The development of such a device was prompted by the fact astronomers realized that photonic technologies offered advanced functionalities like spectral and spatial filtering with numerous advantages such as small footprints and the possibility to easily thermally stabilize instruments [8]. However, efficient coupling from a telescope into a SMF is extremely difficult because of atmospheric turbulence, which distorts the wavefront of the light destroying the image quality and the mismatch in the pupil illumination between the telescope and a SMF [9, 10]. The photonic lantern presented one approach to being able to utilize diffraction-limited functionality while operating in the seeing-limit.

Early lanterns were constructed by placing a bundle of SMFs into a capillary and tapering them down until the cores of the SMFs had vanished, and only a single multimode core was left. This method was not scalable to systems that required 100's or even 1000's of SMFs as would be required if operating in the visible for example. Two solutions were devised: the first was to exploit integrated photonics [11,12] and the second was to use multi-core fibers [13]. The latter approach realized a fiber which had 120 single-mode cores, each with a diameter of 3.9 μm spaced by 16.9 μm, all within a single cladding of 230 μm in diameter. To achieve the photonic lantern functionality, one end of the fiber was tapered as described above to form the multi-mode input, which could be placed in the focal plane of a telescope. This approach made it relatively simple to route the light to a remote spectrograph. However, it also offered the possibility to imprint fiber Bragg gratings across all fiber cores in order to spectrally filter OH-contaminated parts of the spectrum before injecting into a spectrograph. This is an important application for studying high redshift galaxy spectra, where the spectral features of interest are typically hidden amongst the forest of absorption lines created by the atmosphere in the near-IR (1-2 μm) [13].

## 4. Multicore single-mode IFUs for the high-resolution spectroscopic characterization of exoplanets

To directly image an exoplanet requires the ability to differentiate the faint planet from a sea of atmospheric and quasi-static (optical train induced) speckles. Techniques such as angular differential imaging rely on the fact the planet moves through the field during an observing sequence while the speckles remain fixed to locate the planet. Spectral differential imaging relies on the chromatic properties of speckles and more specifically the fact that a speckle moves outwards from the star in an image as a function of wavelength, while a planet remains fixed.

Recently a more ambitious approach has been proposed to differentiate the speckles from the planet. This relies on collecting the light from the known exoplanet into a fiber behind a coronagraph [10, 14]. The coronagraph is used to suppress the on-axis starlight reducing the photon noise contribution. However, residual wavefront error and defects cause stellar leakage and some of the starlight is still present in the focal plane. Because the star is many orders of magnitude brighter than the planet, even the little leaked light can swamp the planetary signal. To address this issue, light can be collected with a second fiber (or more) away from the planet location so that a spectrum of the star with atmospheric tellurics imprinted can be taken simultaneously. Indeed, using more fibers to collect light throughout the focal plane allows for a more accurate removal of the stellar spectrum from that of the planet as they are vastly different. This technique has never before been attempted on-sky but instruments are being developed to pursue this.

Although currently under development for the characterization of known exoplanets, this technique can be used for blind searches for new exoplanets as well. Given the faint nature of these sources and the long integration times needed this would be extremely time consuming unless an integral field unit, which consists of many fiber elements that are very closely packed is used. This would be an ideal application for multicore fibers and an example core geometry is shown in Fig. 2 below. This fiber and technique will be tested with the SCExAO instrument [15].

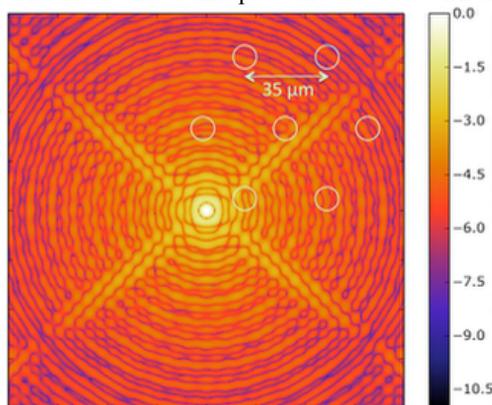

Fig. 2. An image of a star in the focal plane of the Subaru Telescope (logarithmic scale). The overlaid image represents a multicore fiber with 35 μm spacing between cores on a hexagonal grid. The multicore fiber is single mode at 1550 nm.

## 5. Summary

Here we have offered an overview of how multicore fibers are being used in astronomy. Fibers with many cores with a close packing geometry are important for any application that requires spatially resolved spectroscopy of an object. Due to the difficulty of coupling efficiently into SMFs, MMF based multicore fibers will continue to be used. The other key application of these fibers is developing high port count photonic lanterns, which can have matched spectral filters across all ports. As these fibers continue to evolve we will see new instrumentation developments that will no doubt open up new areas of scientific research in astronomy and astrophysics.